# Coulomb potential screening via charge carriers and charge-neutral dipoles/excitons in two-dimensional case

Ke Xiao,[1,2,†,*] Chi-Ming Kan,[2,3,†] Stuart. S. P. Parkin[1] and Xiaodong Cui[2]

[1]*NISE Department, Max Planck Institute of Microstructure Physics, Halle, Germany*

[2]*Department of Physics, The University of Hong Kong, Pokfulam, Hong Kong*

[3]*Department of Chemistry, Hong Kong University of Science and Technology, Clear Water Bay, Hong Kong*

\* Corresponding author: Ke.Xiao@mpi-halle.mpg.de
† These authors contribute equally.

## Abstract

With the shrinking of dimensionality, Coulomb interactions play a distinct role in two-dimensional (2D) semiconductors owing to the reduced dielectric screening in the out-of-plane direction. Apart from dielectric screening, free charge carriers and/or dipoles can also make a non-negligible contribution to Coulomb interaction. While the Thomas-Fermi model is effective in describing charge carrier screening in three dimensions, the extent of screening to two dimensions resulting from charge carriers and charge-neutral dipoles remains quantitatively unclear. Herein, we present an analytical solution based on linear response theory, offering a comprehensive depiction of the Coulomb screened potential in both 2D and 3D systems, where screening effects from both charge carriers and charge-neutral dipoles are addressed. Our work provides a useful and handy tool for directly analysing and evaluating Coulomb interaction strength in atomically thin materials, particularly in the context of electronic and optoelectronic engineering. As a demonstration, we utilized the derived modified Coulomb potential for the exciton system in 2D semiconductors to estimate the exciton binding energy variation arising from the exciton density fluctuation and temperature-dependent exciton polarizability, yielding excellent agreement with the computational and experimental findings.

## I. INTRODUCTION

The emergence of atomically thin two-dimensional (2D) materials not only offers a versatile platform for physical research, but also holds great promise for various applications owing to their intriguing properties. With reduced dimensionality, the Coulomb interaction is greatly enhanced due to reduced dielectric screening and spatial confinement.[1,2] This enhanced Coulomb interaction plays a more significant role in the electronic properties of 2D materials than in their three-dimensional (3D) counterparts, usually determining the characteristic optical and electronic properties of 2D materials. Renowned evidence includes the giant exciton binding energy [3-5], significant renormalization of the electronic bandgap [6,7], Moiré excitons in 2D heterostructures [8-10], and enhanced superconductivity [11,12]. Achieving an effective modification of the Coulomb interaction is crucial for potential applications based on 2D materials.[13-15]

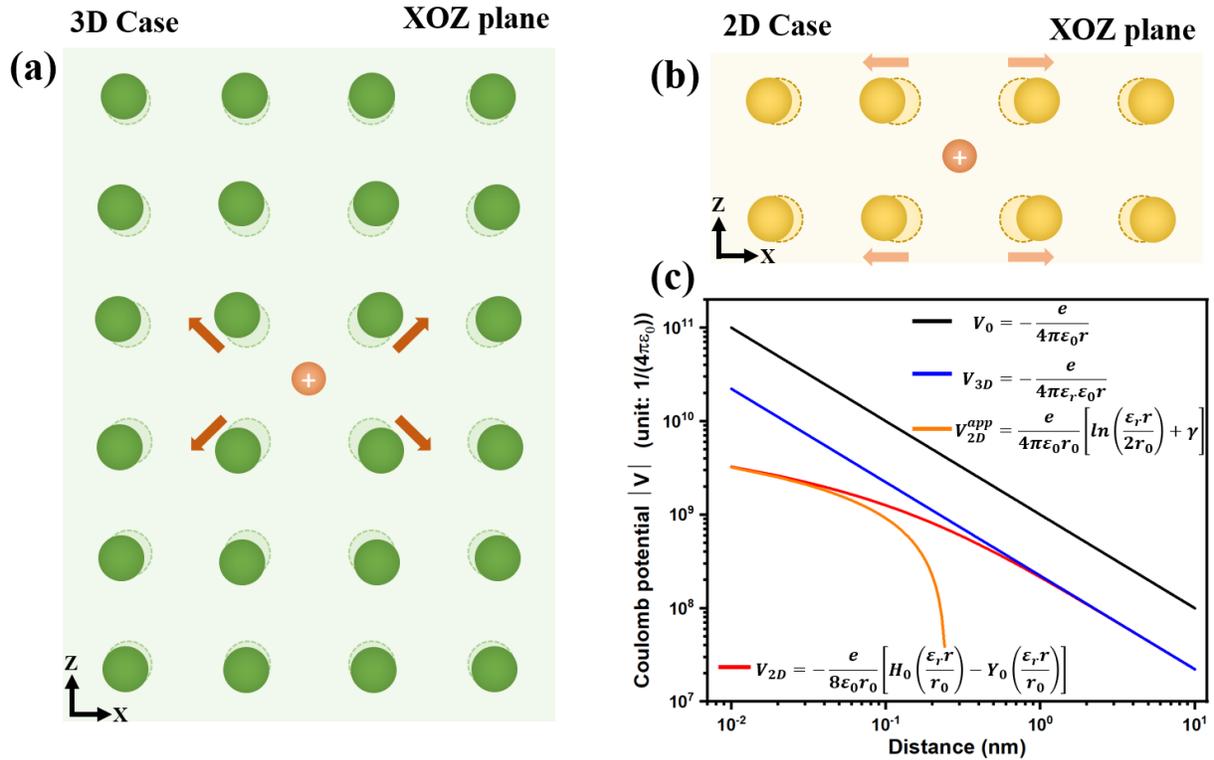

FIG. 1. Dielectric polarization in 3D case (a) and 2D case (b). The dashed (solid) balls represent initial (final) position of ions in crystal lattice. In 2D case, the system is polarizable inside the 2D plane, leading to a nonlocal macroscopic screening. (c) Coulomb potential under different screening situations.

In contrast to the 3D dielectric case, where macroscopic Coulomb screening is well described by a single macroscopic dielectric constant $\varepsilon_r$ (more accurately, a permittivity tensor in the modified Coulomb potential) (Fig.1(a)), macroscopic screening in 2D dielectric cases is highly nonlocal [16-18] and the induced polarization is confined to the 2D plane (Fig.1(b)), resulting in the suppression of dielectric screening in the out-of-plane direction. The Coulomb potential due to this contrasting polarization in the 2D case is widely described in the Rytova-Keldysh form [19], $V_{2D}(\boldsymbol{\rho}, z) = -\frac{e}{8\varepsilon_0 r_0}\left[H_0\left(\frac{\varepsilon_r \rho}{r_0}\right) - Y_0\left(\frac{\varepsilon_r \rho}{r_0}\right)\right]$, where $\boldsymbol{\rho}$, $r_0$ are the 2D spatial coordinates and effective screening lengths, respectively. It displays logarithmic divergence ($V_{2D}^{app} = \frac{e}{4\pi\varepsilon_0 r_0}\left[\ln\left(\frac{\varepsilon_r \rho}{2r_0}\right) + \gamma\right]$) over a short range and is converted to the conventional 3D screened potential ($V_{3D} = \frac{e}{4\pi\varepsilon_r\varepsilon_0 r}$) in the long-range limit (Fig.1(c)). Apart from polarization from the crystal lattice, the Coulomb potential can be further screened when free charge carriers and charge-neutral dipoles exist. Typically, in 3D systems, the screened Coulomb potential arising from charge carriers is well described by the Debye model, Thomas-Fermi model or the Lindhard theory [20]. An exponential damping term is exerted on the long-range Coulomb potential ($V_{3D}^{carrier} = \frac{e}{4\pi\varepsilon_r\varepsilon_0 r}e^{-r/r_D}$), where the $r_D$ is the Debye screening length, making it short range. For 2D systems, although electrostatic doping has been extensively studied in different kinds of 2D materials [21-29] and the screened Coulomb potential due to doped carriers has already been derived and discussed [30,31], the interplay between bandgap renormalization [32-34] and the polaron effect [33,35] makes it difficult to separately resolve the screening effect experimentally. In addition to the contribution from charge carriers, the screening effect of dipoles, which is usually neglected owing to its charge neutrality, could be significant in some scenarios, particularly under a modest dipole density.[36,37] Recently, the screening effect of electric dipoles has been experimentally addressed in exciton system of monolayer transition metal dichalcogenides (TMDs) [38,39]. Inspired by these studies, efforts have been devoted to investigating how electric dipoles contribute to screening Coulomb interactions at the microscopic level. Furthermore, a general and specific quantitative description and comparison of the screened Coulomb potential arising from charge carriers and charge-neutral dipoles are still lacking.

In this letter, we present an analytical solution to describe the Coulomb screened potential in both 2D and 3D cases based on the linear response theory, where screening effects from both charge carriers and charge-neutral dipoles are considered. Additionally, we crosschecked the results using perturbation theory to validate our approach (Supplementary Note1). Using the derived Coulomb potential, we calculated the exciton binding energy as a function of exciton density and estimated the 2D exciton polarizability. These results were consistent with the experimental results. Our work provides a useful tool for analyzing and estimating the strength of the screened Coulomb interactions in atomically thin materials.

## II. DERIVATION OF MODIFIED COULOMB POTENTIAL UNDER SCREENING EFFECT FROM CHARGE CARRIERS AND CHARGE-NEUTRAL DIPOLES:

### A. Charge carrier screening

To incorporate the effect of carrier screening in the 2D case, we introduce a dielectric sheet (with subscript $m$) with negligible thickness encapsulated in a dielectric environment (with subscript $s$) (Fig.2(a)). Initially, the system is in thermodynamic equilibrium with a uniform carrier density ($n_{carrier}(\boldsymbol{\rho}) = cons.$) as the Jellium model.[40] The system is perturbed by introducing a point charge ($e\delta(\boldsymbol{r})$) at the origin and at time $t_0$, described by a time-dependent perturbative term ($H_{1,s}(t) = \begin{cases} 0, & t < t_0 \\ V_I(t), & t > t_0 \end{cases}$). As a result, the carrier density at time t is raised by an induced carrier density defined as $n_{carrier}^{ind}(\boldsymbol{\rho}, t) = n_{carrier}(\boldsymbol{\rho}, t) - n_{carrier}(\boldsymbol{\rho}, 0)$.

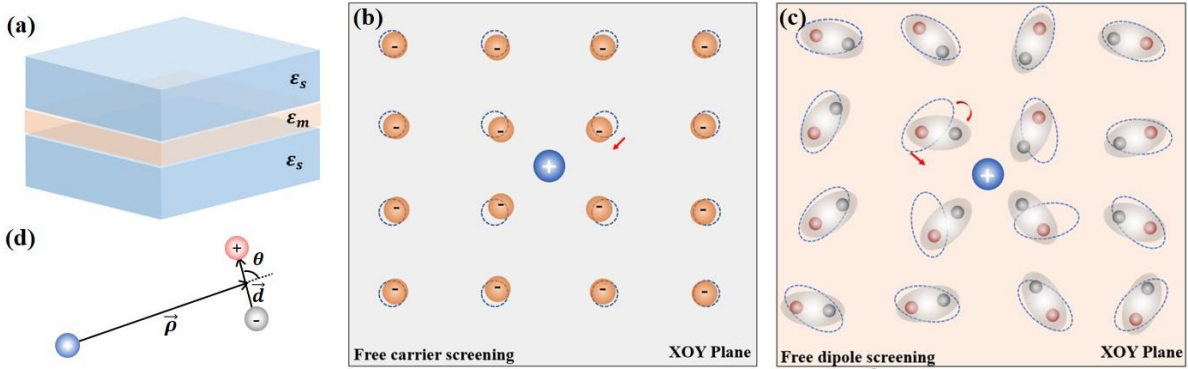

FIG. 2. Schematic representation of a dielectric sheet ($\varepsilon_m$) encapsulated in a dielectric environment ($\varepsilon_s$) (a) and screening effect from charge carriers (b) and charge-neutral dipoles (c).

The screened Coulomb potential ($\varphi(\boldsymbol{r})$) arising from the external point charge at the origin can be expressed by Poisson's equation:

$$\nabla^2 \varphi(\boldsymbol{r}) = -\frac{e}{\varepsilon_0}\delta(\boldsymbol{r}) - \frac{\eta^{ind}(\boldsymbol{r})}{\varepsilon_0} \tag{1}$$

Here, $\varepsilon_0$ is vacuum permittivity, $\eta^{ind}(r) = en^{ind}(r)$ represents the induced charge density given by $en^{ind}(r) = \alpha_s \nabla^2 \varphi(r) + (\alpha_m^{2D} - \alpha_s^{2D})\nabla_\rho^2 \varphi(\rho, z = 0)\delta(z) + en_{carrier}^{ind}(\rho, z = 0)\delta(z)$, of which the final term encodes the screening effect from charge carriers. $\alpha_s$ is the 3D polarizability, $\alpha_s^{2D}$ and $\alpha_m^{2D}$ represent the 2D polarizabilities of the surroundings and dielectric sheet, which are linked to the macroscopic polarization by $\boldsymbol{P}_s = -\alpha_s \nabla\varphi(r)$, $\boldsymbol{P}_s^{2D} = -\alpha_s^{2D}\nabla_\rho\varphi(\rho, z = 0)$, $\boldsymbol{P}_m^{2D} = -\alpha_m^{2D}\nabla_\rho\varphi(\rho, z = 0)$, respectively.

Under the interaction picture, the dynamics of the induced carrier density in the 2D case can be written using linear response theory:

$$n_{carrier}^{ind}(\boldsymbol{\rho}, t) = -\frac{i}{\hbar}\Theta(t - t')\int_{-\infty}^{\infty} dt' \langle [\hat{n}_I(\boldsymbol{\rho}, t), V_I(t')]_- \rangle \tag{2}$$

Where $\Theta(t) = -\lim_{\eta \to 0^+} \frac{1}{2\pi i} \int_{-\infty}^{\infty} d\omega \frac{e^{-i\omega t}}{\omega + i\eta}$, is the Heaviside step function, $V_I(t') = e\varphi(\boldsymbol{\rho}', t')\hat{n}(\boldsymbol{\rho}', t')$ is the potential energy coupled to the carrier density at time $t'$. According to the definition of the two-particle Green's function $\chi_{carrier}(\rho, \rho'; t, t') = -\frac{i}{\hbar}\Theta(t - t')\langle[\hat{n}_I(\boldsymbol{\rho}, t), \hat{n}_I(\boldsymbol{\rho}', t')]_-\rangle$, the induced carrier density can be further written as

$$n_{carrier}^{ind}(\boldsymbol{\rho}, t) = e\int_{-\infty}^{\infty} dt' d\rho' \chi_{carrier}(\boldsymbol{\rho} - \boldsymbol{\rho}'; t - t')\varphi^{ext}(\boldsymbol{\rho}', t') \tag{3}$$

Insert Eq. (3) into Eq. (1) and perform a Fourier Transform (FT) on both sides with the static limit ($\omega \to 0$):

$$(q^2 + k_z^2)\varphi(\boldsymbol{q}, k_z) = -\frac{1}{\varepsilon}\left(e + q^2\frac{\alpha_{total}^{2D}}{2\pi}\int_{-\infty}^{\infty}\varphi(\boldsymbol{q}, k_z)dk_z\right) - \frac{e^2}{2\pi\varepsilon}\chi(\boldsymbol{q})\int_{-\infty}^{\infty}\varphi(\boldsymbol{q}, k_z)dk_z \tag{4}$$

where $\varepsilon = \varepsilon_0 + \alpha_s$, $\alpha_{total}^{2D} = \alpha_m^{2D} - \alpha_s^{2D}$. The Fourier transform of $\chi_{carrier}$ is obtained to be $\chi_{carrier}(\boldsymbol{q}) = \frac{1}{S}\sum_k \frac{f(\epsilon_{k+q}) - f(\epsilon_k)}{\epsilon_{k+q} - \epsilon_k}$, where $f(\epsilon_k)$ represents the Fermi-Dirac distribution with energy $\epsilon_k$ and $S$ is the real space area of the 2D system. After solving the equation, we obtain the modified Coulomb potential rising from the carrier screening in reciprocal space:

$$\varphi(\boldsymbol{q}) = \frac{e}{q(2\varepsilon + q\alpha_{total}^{2D}) - e^2\chi_{carrier}(\boldsymbol{q})} \tag{5}$$

**B.    Dipole screening:**

Unlike charge carriers, electric dipoles ($p = ed$) are charge-neutral as a whole, but have a finite charge-separated vector. Intuitively, dipole screening can contribute to the screening electrostatic potential via a dipole shift and reorientation (Fig.2(c)). It is usually much weaker than carrier screening and is therefore ignored. However, its contribution can be significant in some scenarios, particularly under a moderate dipole density in a 2D system. One example is small polar molecules in solvents, where dipole screening dominates electrostatic interaction [41-43].

Analogous to the previous section, it is imperative to compute the induced charge density stemming from doped free dipoles. We begin by determining the induced dipole density using the linear response theory, in an analogous way to the process for calculating the induced carrier density.

$$n_{dipole}^{ind}(\boldsymbol{\rho}, t, \theta) = ef(\theta) \times \int_{-\infty}^{\infty} dt' d\boldsymbol{\rho}' \chi_{dipole}(\boldsymbol{\rho} - \boldsymbol{\rho}'; t - t') \varphi_{\theta}^{ext}(\boldsymbol{\rho}', t') \tag{6}$$

Here, $\theta$ represents the angle between the relative distance $\boldsymbol{\rho}'$ and the dipole orientation $\boldsymbol{d}$ (Fig.2(d)), $f(\theta)$ is the ratio parameter for different dipole orientations. The potential energy between the external point charge located at the origin and dipole at position $\boldsymbol{\rho}'$ and time $t'$ with in-plane orientation $\theta$ is given by $e\varphi_{\theta}^{ext}(\boldsymbol{\rho}', t') = -e\varphi^{ext}\left(\boldsymbol{\rho}' - \frac{\boldsymbol{d}_{\theta}}{2}, t'\right) + e\varphi^{ext}\left(\boldsymbol{\rho}' + \frac{\boldsymbol{d}_{\theta}}{2}, t'\right)$

Assuming a highly randomized dipole system ($f(\theta) = cons.$) in a semiclassical way, we can write the induced charge density as

$$\eta_{dipole}^{ind}(\boldsymbol{\rho}, z = 0) = e \sum_{\theta} \left( n_{dipole}^{ind}\left(\boldsymbol{\rho} - \frac{\boldsymbol{d}_{\theta}}{2}, z = 0\right) - n_{dipole}^{ind}\left(\boldsymbol{\rho} - \frac{\boldsymbol{d}_{\theta}}{2}, z = 0\right) \right) \tag{7}$$

Inserting Eq. (7) into Eq. (1) and solving the FT of Poisson's equation in a similar manner, we can obtain the modified Coulomb potential due to dipole screening in reciprocal space:

$$\varphi(\boldsymbol{q}) = \frac{e}{q(2\varepsilon + q\alpha_{total}^{2D}) - 2e^2 \chi_{dipole}(\boldsymbol{q})(1 - J_0(qd))} \tag{8}$$

where $\chi_{dipole}(\boldsymbol{q}) = \frac{1}{S} \sum_k \frac{f(\epsilon_{k+q}) - f(\epsilon_k)}{\epsilon_{k+q} - \epsilon_k}$, $f(\epsilon_k)$ here represents the Bose-Einstein statistic due to dipole being bosonic entities, $J_0$ is the zeroth order Bessel function. Note that $2(1 - J_0(qd)) = \sum_{\theta} 4\sin^2(\boldsymbol{q} \cdot \frac{\boldsymbol{d}_{\theta}}{2})$, which originates from the FT of the electric dipole. (more details are provided in Supplementary Note 1).

| | 2D | 3D |
|---|---|---|
| No screening | $\dfrac{e}{q(2\varepsilon + \alpha_{2D}q)}$ (a) | $\dfrac{e}{\varepsilon q^2}$ |
| Carrier screening | $\dfrac{e}{q(2\varepsilon + q\alpha_{2D}) - e^2\chi(q)}$ (b) | $\dfrac{e}{\varepsilon q^2 - e^2\chi(q)}$ (c) |
| Dipole screening | $\dfrac{e}{q(2\varepsilon + q\alpha_{2D}) - 2e^2\chi(q)(1 - J_0(qd))}$ | $\dfrac{e}{\varepsilon q^2 - 2e^2\chi(q)(1 - j_0(qd))}$ (d) |

TABLE I. Coulomb potential in 2D and 3D systems without screening, with carrier screening and dipole screening, respectively. (a) Ref [2] (b) Ref [30, 31] (c) Lindhard function (d) the long wavelength approximation ($q \to 0$) result was discussed in ref [44].

The screened potential in the 3D system was calculated analogously, and the results are summarized in Table I for comparison. The construction of $\chi_{carrier}(q)$ and $\chi_{dipole}(q)$ indicates that the Coulomb potential remains radially symmetric because homogeneous particles, including charge carriers or charge-neutral dipoles, do not break the radial symmetry in the 2D plane. When considering both the charge carrier and dipole screening effects simultaneously and ignoring the interactions between the carriers and dipoles, a more general modified Coulomb potential can be written as

$$\varphi(q) = \frac{e}{q(2\varepsilon + q\alpha_{total}^{2D}) - e^2\chi_{carrier}(q) - 2e^2\chi_{dipole}(q)(1 - J_0(qd))} \tag{9}$$

Note that $\chi_{carrier}(q)$ and $\chi_{dipole}(q)$ share a similar formula because the two-particle Green's function is applicable to both fermions (carriers) and bosons (dipoles). (Supplementary Note 2).

### III. DISCUSSION OF THE MODIFIED COULOMB POTENTIAL:

In 3D systems, the macroscopic screening effect can be simply described as effective ($\varepsilon_r$) in reciprocal space ($\varphi^{3D}(r)\dfrac{e}{4\pi\varepsilon_r\varepsilon_0}\dfrac{1}{r}$, $\varphi^{3D}(q) = \dfrac{e}{\varepsilon_r\varepsilon_0}\dfrac{1}{q}$). While in 2D system, it is much more complicated. The highly nonlocal effective dielectric constant is intrinsically $q$-dependent and is defined as $\varepsilon_{2D}(q) = \varepsilon_r + \dfrac{\alpha_{2D}q}{2\varepsilon_0}$.[2] Hence, the effective dielectric constant in the modified 2D Coulomb potential owing to carrier and dipole screening effects is defined analogously as $\varepsilon_{carrier}^{2D,eff} = \varepsilon_r + \dfrac{\alpha_{2D}q}{2\varepsilon_0} - \dfrac{e^2\chi(q)}{2q\varepsilon_0}$ and $\varepsilon_{dipole}^{2D,eff} = \varepsilon_r + \dfrac{\alpha_{2D}q}{2\varepsilon_0} - \dfrac{e^2\chi(q)(1-J_0(qd))}{q\varepsilon_0}$. Notably, the dipole-modified Coulomb potential depends on both the doping density ($n$) and dipole size ($d$). In the following calculations, the dipole size was set to be 1.1 nm as a practical example to evaluate the modified Coulomb potential. Fig.3 (a) and (b) show the 2D map of $q$-dependent effective dielectric constant as functions of the carrier and dipole density, respectively. Fig.3(c) and 3(d) summarize the effective dielectric constant and the effective Coulomb potential in reciprocal space as functions of the scattering wave vector $q$ and real spatial distance $\rho$ at a typical carrier density ($10^{11}$ cm$^{-2}$) and dipole density ($10^{11}$ cm$^{-2}$) for comparison. The effective dielectric constant increases with increasing free carriers (Fig.3(a)) or dipole density (Fig.3(b)), indicating that the 2D Coulomb potential is strongly influenced by the presence of carriers and dipoles.

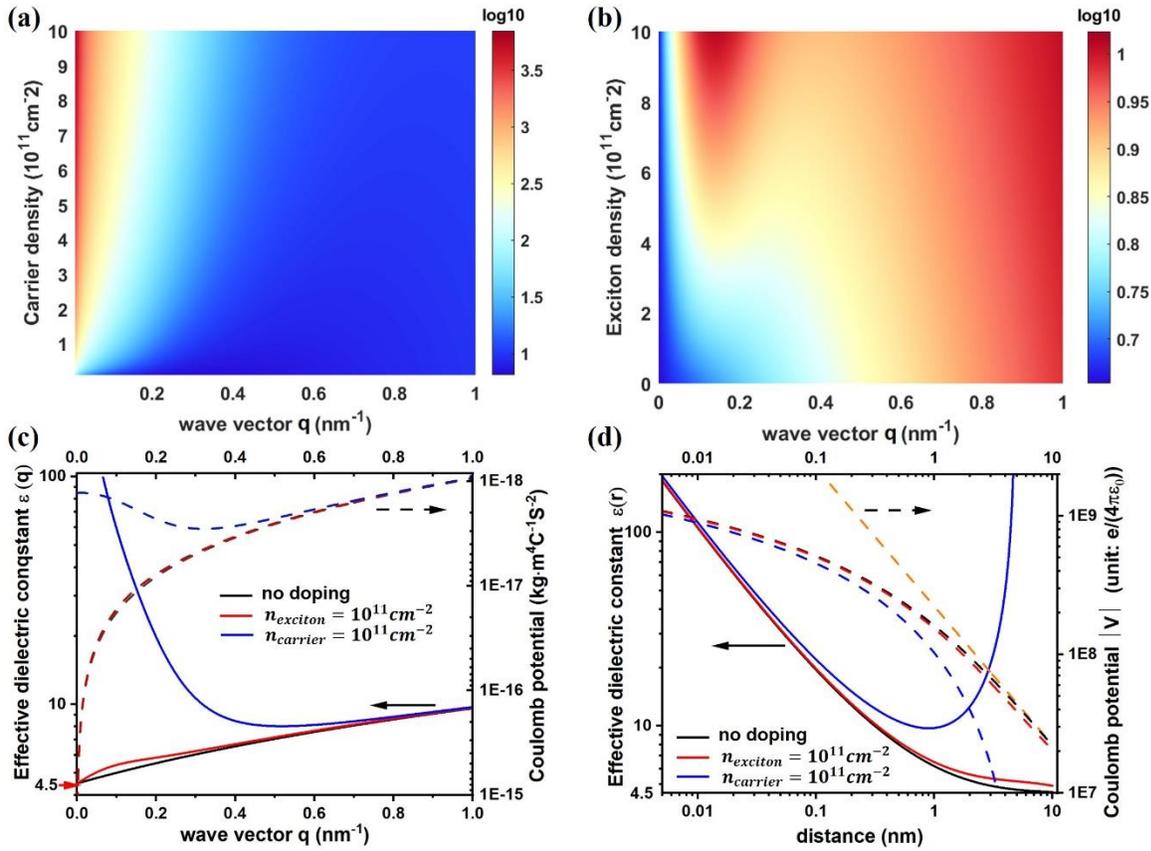

FIG. 3. 2D maps of effective dielectric constant as functions of wave vector, carrier density (a) and dipole density (b). (c) Typical screened $q$-dependent effective dielectric constant (solid line) and Coulomb potential (dashed line) at $n_{dipole} = 10^{11} cm^{-2}$ (red) and $n_{carrier} = 10^{11} cm^{-2}$ (blue), in comparison to the 2D Coulomb potential $\varphi(q) = \frac{e}{q(2\varepsilon + q\alpha_{2D})}$ (black). (d) Typical screened $\rho$-dependent effective dielectric constant (solid line) and Coulomb potential (dashed line) at $n_{exciton} = 10^{11} cm^{-2}$ (red) and $n_{carrier} = 10^{11} cm^{-2}$ (blue), in comparison to the 2D Rytova-Keldysh potential $V_{2D}(\rho, z = 0) = \frac{e^2}{8\varepsilon_0 r_0}\left[H_0\left(\frac{\varepsilon_r \rho}{r_0}\right) - Y_0\left(\frac{\varepsilon_r \rho}{r_0}\right)\right]$ (black) and $V_{3D}(r) = \frac{e}{4\pi\kappa\varepsilon_0 r}$ (orange).

## A. Screened Coulomb potential in momentum space

Under long wavelength approximation ($q \to 0$), for carrier screening, the $\varepsilon_{carrier}^{2D,eff}$ displays a singularity at $q = 0$ (blue solid line), while the effective Coulomb potential remains a finite value, which is similar to the behaviour of the Thomas Fermi screening effect in the 3D case ($\varepsilon_{carrier}^{3D,eff} = \varepsilon_r - \frac{e^2 \chi(q)}{\varepsilon_0 q^2}$). This is because, under the approximation of long wavelength limits, $\chi(q) = -\partial n/\partial \mu$ is noncontingent on the wave vector $\boldsymbol{q}$ (Fig S.4). In contrast to the carrier screening effect, the screened Coulomb potential (red dashed line in Fig.3(c)) originating from dipole screening exhibits the opposite behaviour. It shows a surprisingly similar tendency to that of the vacuum 2D Coulomb potential (black dashed line), both diverging at $q = 0$. When comparing the effective dielectric constant of dipole screening ($\varepsilon_{dipole}^{2D,eff}$, red solid line in Fig.3(c)) to that without carriers or dipoles ($\varepsilon_{2D}$, black solid line in Fig.3(c)), the exciton screening effect is found to be significant at the range $q = 0$~$0.4$ nm$^{-1}$, related to the setting dipole size $d = 1.1$ nm.

While taking short wavelength and low-density approximation ($q \gg k$), $\chi(q)$ can be approximated as $2nm/\hbar^2 q^2$, which is dependent on $q$ as illustrated in FigS.4. It approaches 0 as $q$ increases, implying that the screened Coulomb potential owing to either the carrier (blue dashed line) or dipole (red dashed line) exhibits an asymptotical behaviour towards the unscreened 2D Coulomb potential (black dashed line), as shown in Fig.3(c).

## B. Screened Coulomb potential in real space:

Owing to the charge-neutral nature of excitons/dipoles, their screening effect is much weaker than that of charge-carrier screening. Consequently, the contribution of dipole screening to the modified Coulomb potential is insignificant, particularly at short separations. Notably, the effective dielectric constant shows a non-monotonic dependence on the separation. The initial decrease results from the polarization of the 2D lattice, similar to the 2D Coulomb potential without carriers and dipoles (black solid line in Fig.3(d)). The following increase of the dielectric constant is similar to the carrier screening effect in the 3D case, as described by the Thomas-Fermi potential ($V_{3D}^{carrier}(\boldsymbol{r}) = \frac{e}{4\pi\varepsilon_r\varepsilon_0 r} e^{-\frac{r}{r_0}}$, where the $r$-dependent effective dielectric constant shows an exponential increase

($\varepsilon(r) = \varepsilon_r e^{\frac{r}{r_0}}$). This increase in the effective dielectric constant at large separations indicates the short-range and localized nature of the modified Coulomb potential due to carrier screening.

### D. Dynamic screening

In highly doped systems, the collective motion of the doped carriers or dipoles becomes significant, leading to a transient electric field. Consequently, the screening process occurs almost instantaneously, rendering static screening inadequate and necessitating consideration of screening as a dynamic process. One example of this dynamic nature is the self-energy of electrons or holes, which is predominantly influenced by high-frequency responses, indicative of dynamic screening effects that ultimately result in bandgap renormalization. [32,45] In our calculations, the dynamical effect modifies a frequency-dependent screened Coulomb potential by:

$$\varphi(q,\omega) = \frac{e}{q(2\varepsilon(\omega) + q\alpha_{total}^{2D}(\omega)) - e^2\chi_{carrier}(q,\omega) - 2e^2\chi_{dipole}(q,\omega)(1 - J_0(qd))} \quad (10)$$

where $\chi(q,\omega) = \frac{1}{S}\sum_k \lim_{\eta \to 0^+} \frac{f(\epsilon_{k+q}) - f(\epsilon_k)}{\hbar\omega + \epsilon_{k+q} - \epsilon_k + i\hbar\eta}$ becomes frequency-dependent.

## IV. APPLICATION OF THE MODIFIED COULOMB POTENTIAL:

An exciton is a bound quasi-particle consisting of an electron and a hole with a characteristic separation, called the exciton Bohr radius. Unlike regular dipoles, ground-state excitons do not possess permanent dipole moments owing to the s-type exciton envelope function. Nevertheless, excitons can be approximated as the composition of instantaneous dipoles with complete orientations, or in other words, the statistical average of homogeneous dipoles with different orientations, as depicted in the inset of Fig.4(a). Under this simple approximation, we do not distinguish the screening effect between randomly distributed dipoles and excitons in the following text. Owing to the large binding energies of 2D materials, excitons are stable even at room temperature. Meanwhile, the 2D nature also yields a sizeable Bohr radius, usually around 1 nm for the ground states in monolayer TMDs. Such an exciton system provides an unprecedented platform for investigating the dipole-screened Coulomb potential as a function of exciton density in the 2D case.

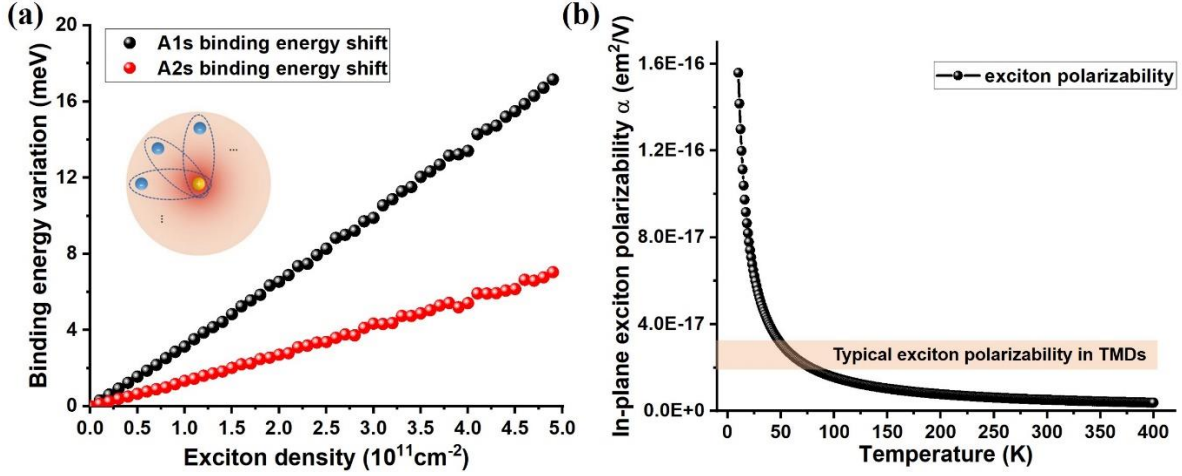

FIG 4. (a) Binding energy shifts of *1s* state and *2s* states of excitons in monolayer TMD due to the exciton screening effect as a function of exciton density. (b) Temperature dependent exciton polarizability with effective dielectric constant ($\varepsilon_r = 4.5$), the orange region shows the typical exciton polarizability ($\sim 2.5 \times 10^{-17} em^2/V$) in monolayer TMDs.

## A. Estimation of exciton binding energy variation as a function of exciton density

As shown in Fig.3(c), the potential modification mainly occurs in the range of 0~0.4 nm$^{-1}$. Given that the effective Bohr radius of ground-state excitons in monolayer TMDs is approximately 1 nm, the screening effect of the exciton population can be significant. Using the dipole screening formula derived in Table I, one can evaluate the change in the Coulomb potential via a perturbative term to the initial Coulomb potential under low or medium exciton densities. Specifically, the modification in the Coulomb potential between exciton-populated and zero-exciton systems, $\Delta V(q)$ can be approximated as $\Delta V(q) \approx \frac{2e^3 \chi(q)(1-J_0(qd))}{q^2(2\varepsilon+q\alpha_{2D})^2}$. This allows us to calculate the exciton binding energy variation of the Rydberg series of excitons (1s, 2s...) using perturbation theory, expressed as

$$\langle \varphi_{ns}(r)|V(r)|\varphi_{ns}(r)\rangle = \frac{1}{(2\pi)^4} \iint \varphi_{ns}^*(k_1)\varphi_{ns}(k_2)\Delta V(k_1-k_2)dk_1 dk_2 \qquad (11)$$

To perform this calculation, we chose the Rydberg exciton wave function from the 2D hydrogen atom model: $\psi_{1s}(k) = 2\sqrt{2\pi}\frac{a_{B,1s}}{[1+a_{B,1s}^2 k^2]^{3/2}}$ and $\psi_{2s}(k) = 2\sqrt{6\pi}\frac{a_{B,2s}}{[1+a_{B,2s}^2 k^2]^{3/2}}\frac{a_{B,2s}^2 k^2-1}{a_{B,2s}^2 k^2+1}$. It should be noted that both the ground state (1s) and the first excited state (2s) exciton wave functions of monolayer TMDs are mainly located in the range of 0~2 nm$^{-1}$, which implies that the modified Coulomb potential due to exciton screening can effectively tune the exciton binding energy.

We performed numerical calculations using the Monte Carlo method, and the results are summarized in Fig.4(b). The binding energies of 1s and 2s excitons increase almost linearly in the low and medium exciton density regimes ($10^{10}$–$10^{12}$ cm$^{-2}$), which is consistent with experimental reports.[39] The calculated binding energy shift is slightly overestimated compared to the experimental results, which may result from the difference between the actual wave functions in 2D materials and the wave functions of the 2D hydrogen atom model used in our numerical calculation.

**B. Estimation of exciton polarizability in 2D materials:**

Assuming excitons to be rigid objects with modest and uniform polarizability ($\alpha_{2D}^{ex}$), the modified Coulomb potential takes the form of $\varphi(q, z=0) = \frac{e}{q(2\varepsilon_0 + \alpha_{2D}q) + \alpha_{2D}^{ex}nq^2}$, [39] which means that the exciton screened part ($\alpha_{2D}^{ex}nq^2$) is linearly dependent on the exciton density $n$ in a classical way. Under the long-wavelength approximation ($q \to 0$), the screened Coulomb potential due to excitons (Eq. (8)) can be simplified as $\varphi(q) = \frac{e}{q(2\varepsilon_0 + \alpha_{2D}q) + nq^2e^2d^2/2k_BT}$, where $n, k_B, T$ are the exciton density, Boltzmann constant, and effective exciton temperature, respectively. By comparing these two modified Coulomb potentials, it can be concluded that the exciton polarizability is a function of the effective exciton Bohr radius ($d$) and effective exciton temperature in a simplified formula of $\alpha_{2D}^{ex} = \frac{e^2d^2}{2k_BT}$. With increasing Bohr radius, the exciton polarizability experiences quadratic growth because it contributes more to the polarization per dipole. On the other hand, as the temperature increases, the disorder steps in and lead to a decrease in exciton polarizability. Considering the environmental dielectric property ($\varepsilon_r = 4.5$), the temperature-dependent exciton polarizability is plotted in Fig.4b. The typical exciton polarizability in monolayer TMDs is measured to be approximately $2.5 \times 10^{17}$ em$^2$/V using the in-plane Stark effect [46-49] or density-dependent photoluminescence measurement at 10 K [39], which is slightly smaller than the theoretical estimation ($\sim 7.8 \times 10^{17} em^2/V$). This overestimation may result from the fact that the effective exciton temperature is usually several tens of Kelvins higher than the lattice temperature (15 K) at non-resonant excitation.[50]

**V. DISCUSSION AND SUMMARY**

Although estimating the exciton binding energy with elevated exciton density using the modified Coulomb potential has proven successful, the derived formula fails to accurately align the estimation of exciton binding energy as a function of carrier density. Although some researchers attribute this discrepancy to the limitation of static screening [31], we contend that its ineffectiveness stems from two key factors. First, the exciton resonance peak comprises various contributing factors including the exciton resonance energy, electronic bandgap, exciton binding energy, exciton-carrier interaction, and exciton-exciton interaction. Particularly in the context of carrier doping, the term $E_{exciton-carrier}$ is more aptly described by the Fermi-polaron picture, which arises from the collective interaction between excitons and the Fermi sea. [23,24] This complexity undermines the simplistic view of excitons solely as a two-particle (exciton) or three-particle system (trion), thereby rendering the application of static screening approximation ineffective in this scenario. Second, it is essential to recognize that both static and dynamic screening approaches are grounded in linear response theory or perturbation theory. The validity of the screened Coulomb potential depends on whether the conditions fit in the perturbation terms and whether static or dynamic screening methodologies are employed. Consequently, the screened Coulomb potential was not applicable at high carrier densities. Nevertheless, for exciton doping at approximately $5 \times 10^{11} cm^{-2}$, the screened term can still be considered a perturbative term, as depicted in Fig.S5.

Furthermore, experimental studies on exciton doping have revealed that the exciton screening effect from mutual polarization outweighs the direct exciton-exciton interaction by a factor of five.[39] Consequently, the screening effect of neutral excitons significantly influences the binding energy, as evidenced by the energy variation between the ground state and the first excited-state exciton.

In summary, we presented an effective model of the Coulomb potential incorporating the screening effect from charge carriers and charge-neutral excitons/dipoles in 2D systems. Our methodology is based on the framework of linear response theory, which is inherently a perturbative approach. The model shows that excitons and charge carriers can modify the 2D Coulomb potential. In conclusion, the modified Coulomb potential offers a simple and direct way to evaluate the strength of Coulomb interactions and can serve as a design tool for Coulomb potential engineering in 2D systems.

**ACKNOWLEDGMENTS**

This work was supported by the National Key R&D Program of China (2020YFA0309600), the Guangdong-Hong Kong Joint Laboratory of Quantum Matter, and the University Grants Committees/Research Grants Council of Hong Kong SAR (AoE/P-701/20, 17300520). The authors thank Prof. Wang Yao, Dr. Jianju Tang and Dr. Yicheng Guan for their fruitful discussions.## Reference

[1]     A. Chernikov, T. C. Berkelbach, H. M. Hill, A. Rigosi, Y. L. Li, O. B. Aslan, D. R. Reichman, M. S. Hybertsen, and T. F. Heinz, Phys Rev Lett **113** (2014).
[2]     P. Cudazzo, I. V. Tokatly, and A. Rubio, Phys Rev B **84** (2011).
[3]     B. R. Zhu, X. Chen, and X. D. Cui, Sci Rep-Uk **5** (2015).

[4] Z. L. Ye, T. Cao, K. O'Brien, H. Y. Zhu, X. B. Yin, Y. Wang, S. G. Louie, and X. Zhang, Nature **513**, 214 (2014).
[5] C. D. Zhang, A. Johnson, C. L. Hsu, L. J. Li, and C. K. Shih, Nano Lett **14**, 2443 (2014).
[6] M. M. Ugeda *et al.*, Nat Mater **13**, 1091 (2014).
[7] K. F. Mak, C. Lee, J. Hone, J. Shan, and T. F. Heinz, Phys Rev Lett **105** (2010).
[8] C. H. Jin *et al.*, Nature **569**, E7 (2019).
[9] K. Tran *et al.*, Nature **567**, 71 (2019).
[10] K. L. Seyler, P. Rivera, H. Y. Yu, N. P. Wilson, E. L. Ray, D. G. Mandrus, J. Q. Yan, W. Yao, and X. D. Xu, Nature **567**, 66 (2019).
[11] E. Navarro-Moratalla *et al.*, Nat Commun **7** (2016).
[12] J. D. Zhou *et al.*, Adv Mater **29** (2017).
[13] Y. Xu, S. Liu, D. A. Rhodes, K. Watanabe, T. Taniguchi, J. Hone, V. Elser, K. F. Mak, and J. Shan, Nature **587**, 214 (2020).
[14] Y. H. Tang *et al.*, Nature **579**, 353 (2020).
[15] Y. Bai, Y. Li, S. Liu, Y. Guo, J. Pack, J. Wang, C. R. Dean, J. Hone, and X. Zhu, Nano Lett **23**, 11621 (2023).
[16] D. Y. Qiu, F. H. da Jornada, and S. G. Louie, Phys Rev B **93** (2016).
[17] F. Hüser, T. Olsen, and K. S. Thygesen, Phys Rev B **88** (2013).
[18] K. Noori, N. L. Q. Cheng, F. Y. Xuan, and S. Y. Quek, 2d Mater **6** (2019).
[19] L. V. Keldysh, Jetp Lett+ **29**, 658 (1979).
[20] N. W. M. Aschroft, N. D., *Solid state physics* (Cengage Learning, 2022).
[21] J. S. Ross *et al.*, Nat Commun **4** (2013).
[22] E. Liu, J. van Baren, C. T. Liang, T. Taniguchi, K. Watanabe, N. M. Gabor, Y. C. Chang, and C. H. Lui, Phys Rev Lett **124** (2020).
[23] E. F. Liu, J. van Baren, Z. G. Lu, T. Taniguchi, K. Watanabe, D. Smirnov, Y. C. Chang, and C. H. Lui, Nat Commun **12** (2021).
[24] K. Xiao, T. F. Yan, Q. Y. Liu, S. Y. Yang, C. M. Kan, R. H. Duan, Z. Liu, and X. D. Cui, J Phys Chem Lett **12**, 2555 (2021).
[25] M. H. He *et al.*, Nat Commun **11** (2020).
[26] M. Sidler, P. Back, O. Cotlet, A. Srivastava, T. Fink, M. Kroner, E. Demler, and A. Imamoglu, Nat Phys **13**, 255 (2017).
[27] K. F. Mak, K. L. He, C. Lee, G. H. Lee, J. Hone, T. F. Heinz, and J. Shan, Nat Mater **12**, 207 (2013).
[28] Y. Cao, V. Fatemi, S. Fang, K. Watanabe, T. Taniguchi, E. Kaxiras, and P. Jarillo-Herrero, Nature **556**, 43 (2018).
[29] K. A. Parendo, K. H. S. B. Tan, A. Bhattacharya, M. Eblen-Zayas, N. E. Staley, and A. M. Goldman, Phys Rev Lett **95** (2005).
[30] N. S. Rytova, 2018), p. arXiv:1806.00976.
[31] M. M. Glazov and A. Chernikov, Phys Status Solidi B **255** (2018).
[32] S. S. Ataei and A. Sadeghi, Phys Rev B **104** (2021).
[33] M. G. Kang, S. W. Jung, W. J. Shin, Y. Sohn, S. H. Ryu, T. K. Kim, M. Hoesch, and K. S. Kim, Nat Mater **17**, 676 (2018).
[34] Z. Zhang *et al.*, Acs Nano **13**, 13486 (2019).
[35] D. K. Efimkin, E. K. Laird, J. Levinsen, M. M. Parish, and A. H. MacDonald, Phys Rev B **103** (2021).
[36] C. Neidel *et al.*, Phys Rev Lett **111** (2013).
[37] J. G. Gay, Phys Rev B **4**, 2567 (1971).
[38] P. D. Cunningham, A. T. Hanbicki, K. M. McCreary, and B. T. Jonker, Acs Nano **11**, 12601 (2017).
[39] K. Xiao *et al.*, 2023), p. arXiv:2308.14362.
[40] R. I. G. Hughes, Perspectives on Science **14**, 457 (2006).
[41] D. Beglov and B. Roux, J Phys Chem B **101**, 7821 (1997).
[42] M. V. Fedorov and A. A. Kornyshev, Chem Rev **114**, 2978 (2014).
[43] D. J. Lockhart and P. S. Kim, Science **260**, 198 (1993).
[44] J. H. Collet and T. Amand, Solid State Commun **52**, 53 (1984).
[45] B. Scharf, D. V. Tuan, I. Zutic, and H. Dery, J Phys-Condens Mat **31** (2019).
[46] M. Massicotte *et al.*, Nat Commun **9** (2018).
[47] T. G. Pedersen, Phys Rev B **94** (2016).

[48] B. R. Zhu, K. Xiao, S. Y. Yang, K. Watanabe, T. Taniguchi, and X. D. Cui, Phys Rev Lett **131** (2023).
[49] J. Pu, K. Matsuki, L. Q. Chu, Y. Kobayashi, S. Sasaki, Y. Miyata, G. Eda, and T. Takenobu, Acs Nano **13**, 9218 (2019).
[50] K. D. Xiao, R.; Liu, Z.; Watanabe, K.; Taniguchi, T.; Yao, W.; Cui, X., Natural Sciences **3**, e20220035 (2023).